\documentclass[useAMS]{mn2e}

\usepackage{graphicx}
\usepackage{times}

\def\chandra{{\it Chandra~}}

\def\integral{{\it INTEGRAL~}}

\def\1811{{PSR J1811-1925}}

\begin{document}

\title{INTEGRAL observations of PSR J1811$-$1925 and its associated Pulsar
Wind Nebula}
\author[A.J.~Dean et al.]{A.J.~Dean,$^{1}$ A.~De~Rosa,$^{2}$ V.A.~McBride,$%
^{1}$ R.~Landi,$^{3}$ A.B.~Hill,$^{1}$ L.~Bassani,$^{3}$ \newauthor %
A.~Bazzano,$^{2}$ A.J.~Bird,$^{1}$ P.~Ubertini,$^{2}$ \\
$^1$School of Physics and Astronomy, University of Southampton, Highfield,
SO17 1BJ, United Kingdom \\
$^2$INAF/IASF, via del Fosso del Cavaliere 100. Roma, 00113, Italy \\
$^3$INAF/IASF, via P. Gobetti 101, Bologna, 40129, Italy}
\maketitle

\begin{abstract}
We present spectral measurements made in the soft (20--100\,keV) gamma-ray
band of the region containing the composite supernova remnant G11.2-0.3 and
its associated pulsar \hbox{PSR~J1811$-$1925}. Analysis of INTEGRAL/IBIS data allows
characterisation of the system above 10\,keV. The IBIS spectrum is best
fitted by a power law having photon index $\Gamma =$ 1.8 $_{-0.3}^{+0.4}$
and a 20--100\,keV flux of 1.5 $\times $10$^{-11}$ erg cm$^{-2}$ s$^{-1}$.
Analysis of archival \textit{Chandra} data  over different energy bands
rules out the supernova shell as the site of the soft gamma-ray emission
while broad band (1--200\,keV) spectral analysis strongly indicates that the
INTEGRAL/IBIS photons originate in the central zone of the system which
contains both the pulsar and its nebula. The composite X-ray and soft gamma-ray spectrum 
indicates that the pulsar provides around half 
of the emission seen in the soft gamma-ray domain; its spectrum is hard with  
no sign of a cut off up to at least 80\,keV. The other half of the emission above 10\,keV comes from the PWN;
with a $\Gamma$=1.7 its spectrum is softer than that of the pulsar.  From the IBIS/ISGRI mosaics we are able
to derive 2$\sigma $ upper limits for the 20--100\,keV flux from the
location of the nearby TeV source \hbox{HESS~J1809$-$193} to be 4.8 $\times $10$^{-12}
$ erg cm$^{-2}$ s$^{-1}$. We have also examined the likelihood of an
association between \hbox{PSR~J1811$-$1925} and \hbox{HESS~J1809$-$193}. Although PSR
J1811$-$1925 is the most energetic pulsar in the region, the only one detected
above 10 keV and thus a possible source of energy to fuel the TeV fluxes,
there is no morphological evidence to support this pairing, making it an
unlikely counterpart.
\end{abstract}


\section{Introduction}

Recently, \textit{Chandra~} and \textit{XMM-Newton~} X-ray observations have
allowed detailed morphological studies of Pulsar Wind Nebulae (PWN),
resolving complex structures within the general envelope of the plerion,
including toroidal features, axial jets and wisps on arc second angular
scales; for recent reviews see Kaspi et al. (2006) and Gaensler \& Slane
(2006). The PWN are generally associated with young energetic pulsars with
ages of less than a few tens of thousands of years such as the cases of the
Crab and Vela systems (Weisskopf et al. 2004, Kirsch et al. 2006; Helfand et
al 2001; Pavlov et al. 2003; Mangano et al. 2005 ). It is generally thought
that the radiation mechanism from radio to X-ray wavelengths is due to
synchrotron emission as a result of electrons injected by the pulsar, with
the radio emission due to the relic electrons and X-rays resulting from
freshly injected electrons. The diminishing angular sizes of the spectral images
with increasing photon energy provide strong evidence for synchrotron
cooling in a number of cases. However the \textit{Chandra~} spectral images
show that the separate morphological regions within the overall nebular
configuration, such as a jet, counter jet, torus, extended PWN, moving hot
spots and so on are characterised by distinct spectral indices in the X-ray
domain. This indicates that the transport of the energetic electrons is not
simply ``isotropically outwards'' from a central source, but can be the
result of collimated beams, sometimes containing moving blobs with
considerably harder spectra than the surrounding emission structures (e.g.
Mori et al. 2004 for the case of the Crab nebula). 

The study of these
objects in the soft gamma-ray domain, adjacent to the X-ray band, can
provide key information to disentangle the mechanisms active in different
emitting regions. In view of this, the IBIS gamma-ray imager on board
INTEGRAL is a powerful tool for the study of PWN systems, allowing source
detection above 20\thinspace keV with sensitivity of the order of mCrab.
Moreover, the recent survey performed with the HESS (High Energy
Stereoscopic System) telescope array revealed the existence of emission at
TeV energies from the vicinity of a number of pulsar wind nebulae (Aharonian
et al. 2006; 2007). The combined analysis of X-/gamma-ray and Very High
Energy (VHE) emission is of key importance to estimate the relative
contribution of synchrotron and inverse Compton mechanisms and thus to
derive information on the magnetic field of the nebula and the spatial
energy distribution of the accelerated particles.

\section{PSR~J1811$-$1925}

An interesting case to investigate is the 65\thinspace ms pulsar %
\hbox{PSR~J1811$-$1925}, discovered in X-rays by \textit{ASCA~} in the
supernova remnant (SNR) G11.2-0.3 (Torii et al. 1997). It is a composite SNR
showing both an extended shell component and a compact plerionic component.
This discovery strongly suggests a direct association of the pulsar with the
SNR and with a ``guest star'' witnessed by Chinese astronomers in A.D. 386
as proposed by Clark \& Stephenson (1977). The distance of the system is $%
\sim $5\,kpc as inferred from H~I measurements (Becker, Markert \& Donahue
1985; Green et al. 1988). Thanks to X-ray observations performed by \textit{%
Chandra~} (Kaspi et al. 2001) it was possible to localise the pulsar to within a few arc seconds
from the geometric centre of the shell of the SNR (Tam \& Roberts 2003),
thus providing an indication that the system may be young.

A more detailed analysis of the \textit{Chandra~} data combined with VLA
observations (Roberts et al. 2003) demonstrates the separation of the PWN
from the surrounding shell, suggesting that the reverse shock has not yet
reached the PWN, thus implying the system age is no more than $\sim 2000$%
\thinspace yr. The youth of the system was also confirmed using VLA data to
measure the expansion of G11.2-0.3 over a period of 17 yr, which provides a
direct age estimate in the range 960--3400\thinspace yr, which is compatible
with the $\sim1620$\,yr age derived on the basis of the direct association with the
supernova event of A.D. 386 (Tam \& Roberts 2003). However, if it is assumed
that the energy loss from the pulsar is a result of magnetic dipole
radiation losses, with a braking index of n$\sim $3, the spin down age of
the pulsar, P/2$\dot{P}$, would imply that the supernova explosion took
place around 24000 years ago. This apparent age discrepancy can be
explained by assuming that the initial spin period of the pulsar was very
close to its current value, suggesting that $\dot{E}$ (6.4 $\times $ 10$%
^{36} $ erg s$^{-1}$) did not vary significantly since the supernova
explosion. This state of affairs indicates that pulsar spin
characteristics provide poor estimates of the ages of pulsars, and that
rotating neutron stars may become pulsars with longer periods than
previously thought. If this is the case, it is possible to estimate that
PSR~J1811$-$1925 was created with an initial period of $\sim $62 ms (Torii
et al. 1997, Kaspi et al. 2001)

The system \hbox{PSR~J1811$-$1925}/G11.2-0.3 is located near the TeV source \hbox{HESS~J1809$-$193}, recently discovered by the HESS telescope array (Aharonian et al.
2007). Although the pulsar is energetic enough to power the high energy
gamma-rays, the association with the TeV source is not likely since the
pulsar is clearly the companion of G11.2-0.3 and \hbox{HESS~J1809$-$193} is offset
from the supernova remnant, implying that PSR~J1811$-$1925 could not have
produced the TeV source due to its motion. The likelihood of an association
is further questioned by the presence of a second pulsar/PWN system, J1809$-$%
1917 (Kargaltsev \& Pavlov 2007), which is closer to the centroid of the
HESS source and also capable of supplying sufficient energy from the pulsar
to power \hbox{HESS~J1809$-$193}. 

In this paper, we report the first observation of
G11.2-0.3/\hbox{PSR~J1811$-$1925} above 10\thinspace keV by INTEGRAL/IBIS;
using the soft gamma-ray data in conjunction with archival \textit{Chandra~}
observations we establish the nature of the IBIS emission and critically
discuss the possibility of its association with the newly discovered HESS
source.

\section{The INTEGRAL Observation}

\label{integral}

\begin{figure*}
\includegraphics[width=12cm]{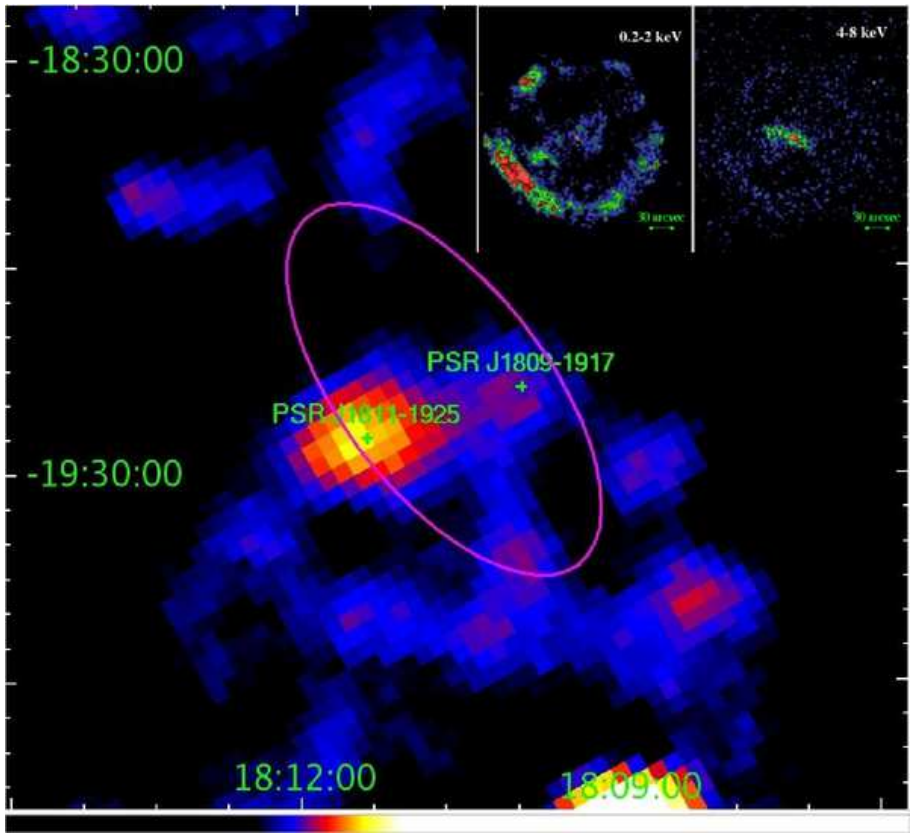}
\caption{\textit{INTEGRAL~} IBIS/ISGRI 18--60\,keV image of the region
around \hbox{PSR~J1811$-$1925}. Note that the image structure around the
position of \hbox{PSR~J1809$-$1917} is well below the detection threshold as
determined by Bird et al. (2007) for the 3rd IBIS/ISGRI catalogue. The fit
ellipse (Aharonian et al. 2007) of the extended HESS source is reported with
a magenta ellipse. In the inset we show the \textit{Chandra~} images ($\sim5 
\arcmin\times5\arcmin$) in 0.2--2\,keV (left) and 4--8\,keV (right).}
\label{chandrahr}
\end{figure*}

The region surrounding \hbox{PSR~J1811$-$1925} has been covered as part of the third 
INTEGRAL IBIS/ISGRI survey (Bird et al. 2007) processing, which
consists of all exposures from the beginning of the mission (November 2002)
up to April 2006. The total exposure on this region is $\sim $3.5\thinspace
Ms. ISGRI images for each available pointing were generated in the 18--60\,keV
band using the ISDC offline scientific analysis software version 5.1
(OSA~5.1; Goldwurm et al. 2003). The individual images were then combined to
produce a mosaic of the entire sky to enhance the detection significance
using the system described in detail by Bird et al. (2004, 2007). Figure~\ref%
{chandrahr} shows the image of the region surrounding \hbox{PSR~J1811$-$1925}%
: a clear excess above the detection threshold as determined by Bird et al.
(2007) is observed with a significance of $\sim $9$\sigma $ at a position
corresponding to R.A.=18$^{\mathrm{h}}$11$^{\mathrm{m}}$ 18$^{\mathrm{s}}$%
.35 and Dec=--19$^{\circ }$ 25$^{\prime }$ 02$^{\prime \prime }$.64
(J2000.0) with a positional uncertainty (90\%) of 2$^{\prime }$.9.  Within
this positional uncertainty the IBIS/ISGRI excess is coincident with
both SNR G11.2-0.3 and its pulsar \hbox{PSR~J1811$-$1925}, despite the fact
that the latter is closer to the centroid of the IBIS emission. INTEGRAL
does not have sufficient angular resolution to discriminate which of these
two (or other components if present) are responsible for the soft gamma-ray
emission, but spectroscopic studies in conjunction with X-ray data can help
to identify or eliminate the various morphological details present in the
region. For this purpose, we obtained the IBIS/ISGRI spectrum following
usual procedures: fluxes were extracted from the location of the source in
thirteen narrow energy bands over the 17--200\,keV range for all available
pointings; a spectral \textit{pha} file was then made by taking the weighted
mean of the light curve obtained in each band. An appropriately re-binned 
\textit{rmf} file was also produced from the standard IBIS spectral response
file to match the \textit{pha} file energy bins. The spectral analysis was
performed using XSPEC v. 11.3.2; quoted errors correspond to 90\% confidence
levels for one interesting parameter ($\Delta \chi ^{2}=2.71$).

A simple power law model provides a good fit to the IBIS/ISGRI data in the
17--200\thinspace keV energy band ($\chi ^{2}/\mathrm{{dof}=16.5/12}$) with
a photon index $\Gamma =1.8_{-0.3}^{+0.4}$ and a 20--100\,keV flux of $%
1.49\times $ 10$^{-11}$\,erg\,cm $^{-2}$\,s$^{-1}$. As there is no \emph{%
INTEGRAL} detection at the location of the TeV object, we can only derive
from the IBIS/ISGRI mosaics $2\sigma $ upper limits to the flux of an
undetected counterpart to be $1.7\times 10^{-12}$ and $3.1\times 10^{-12}$%
\thinspace erg\thinspace cm$^{-2}$ \thinspace s$^{-1}$ in the 20--40 and
40--100\thinspace keV energy bands respectively.

Likewise, no positive excess has been measured by INTEGRAL from the location
of the nearby pulsar/PWN system PSR~J1809$-$1917. The corresponding upper
limits on the soft gamma-ray emission from this system, in the same field
and with the same exposure, are similar to those given for the TeV source
above, and correspond to a $2\sigma $ upper limit on the 20--100\,keV
luminosity of $\sim7\times 10^{33}$\,erg\,s$^{-1}$ for a distance of 3.5\,kpc to \hbox{PSR~J1809$-$1917}.
This soft $\gamma $-ray luminosity implies that \hbox{PSR~J1809$-$1917} is converting
less than $\sim 0.4\%$ of its spin down power into the soft $\gamma $-ray
domain.

\section{\textit{Chandra~} observations}

\label{chandra}

\textit{Chandra~} observed the region around SNR G11.2-0.3 three times
in 2000 and four times in 2003. Roberts et al. (2003) published the analysis
of the observations in 2000 (20 and 15\thinspace ks), performing a spectral
analysis of the different components of the system:  the SNR shell, the
pulsar at the centre and the elongated structure around the pulsar.  This
structure can be interpreted as a jet or Crab like torus seen edge-on and
will be referred from now on as \hbox{PSR~J1811$-$1925} PWN.  Although the X-ray 
images also
show evidence of relativistic dynamic evolution of bright X-ray spots near
the pulsar, the SNR shell, the pulsar and the PWN are the most prominent,
i.e. brightest, components at these energies.
With the specific aim of matching the spectral profile of \ these various
subcomponents with the \integral spectral emission, we reanalysed the
\textit{Chandra~} observations made in 2000 and two of those performed
in 2003, each having an exposure of about 15\thinspace ks. Our main goal
was to gain some insight as to the location of the soft gamma-ray emission
and any relevance to the HESS\ source. In the top inserts in 
Fig.~\ref{chandrahr}
 we show the \textit{Chandra~} images (for observation ID
3911, performed in 2000) of the region around PSR~J1811$-$1925 in two
different energy bands:  0.2--2\thinspace keV and 4--8\thinspace keV. The
images clearly show that the extended PWN and the pulsar are producing the 
hardest
X-rays while the soft (thermal) photons originate in the SNR shell.
 Non-thermal shell emission is also present as discussed by Roberts et 
al. (2003); however,
this component, described by the synchrotron roll-off model, provides a null 
contribution if extrapolated to the IBIS energy range. On the basis of the 
above considerations we can therefore exclude the SNR shell as the site
of any significant emission above 10 keV.

To further investigate the spectral behaviour of the elongated region
containing the jet and/or a torus, we performed a more detailed spectral 
analysis.
We created two spectra: spectrum A was extracted from a box of 
$15\arcsec\times 17%
\arcsec$ around the hard X-ray plume (top right inset of 
Fig.~\ref{chandrahr}), but  excluding the position of the pulsar 
\hbox{PSR~J1811$-$1925}.  Spectrum B  covered the same region as spectrum A, 
but also included the pulsar.
To avoid pile-up contamination  (see Roberts et al. 2003 for details) 
in spectrum B we
excluded the brightest central pixel in the pulsar region. For both spectra 
the
background was taken in one circular region of $20\arcsec$ radius in the
same CCD as the source. To extract spectra we used the \textit{specextract%
} script of CIAO v. 3.3.0.1 for extended sources. In order to check the 
correctness of our procedure, we have also analysed the data not corrected 
for pile-up but applying a cut in energy at 5 keV as below this energy 
pile-up effects are not important.  The fittings results obtained using both 
methods  were the same, implying a good treatment
of any pile-up problem.

\begin{figure}
\centering
\includegraphics[width=0.75\linewidth, angle=-90]{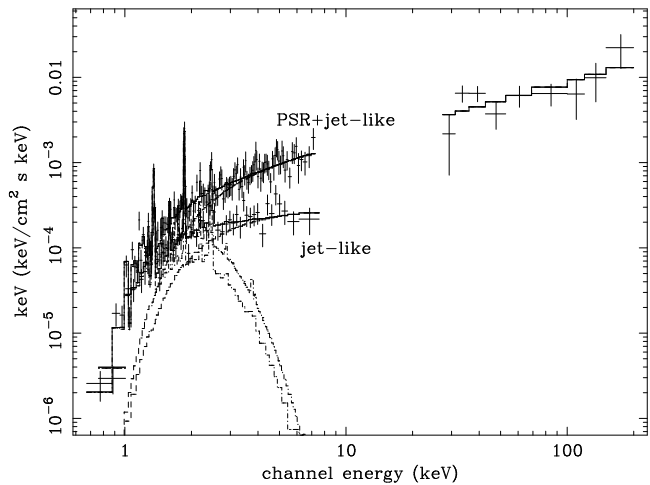}
\caption{Composite  \textit{Chandra~} and \textit{INTEGRAL} spectra. The
two data sets in the \textit{Chandra~} energy range are the spectra extracted
from the jet-like feature (region A in the text) and PSR+jet-like feature
(region B in the text).}
\label{spectrum}
\end{figure}

To model the soft thermal emission that could still contribute to the 
emission in the central region near
the pulsar, we employed the model proposed by Roberts et al. (2003) for both
spectra A and B. This model comprises an absorbed plane-parallel shock 
plasma
model (PSHOCK in XSPEC) with the parameters frozen to the values found in
their analysis (where all regions around the SNR are fitted separately),
with the exception of the absorbing column density $N_{\rm H}$. We add to this
component a simple power-law to fit the PWN alone or the PWN plus the pulsar 
.
The fit we obtained is good for both spectra (A: $\chi ^{2}$/dof=48/57, B: 
$%
\chi ^{2}$/dof=113/127) with photon indices $\Gamma $=1.7$_{-0.4}^{+0.4}$
and $\Gamma $=1.2$_{-0.2}^{+0.2}$ for A and B respectively and 
$N_{\rm H}=(2.4\pm
0.2)\times 10^{22}$\thinspace cm$^{-2}$ in both spectra.  The unabsorbed 
flux
in 1--10\,keV is 3.2$\times $10$^{-12}$ erg cm$^{-2}$ s$^{-1}$ and 6.7$\times 
$%
10$^{-12}$ erg cm$^{-2}$ s$^{-1}$ for A and B respectively. \ Performing
similar analysis of the other \textit{Chandra }observations made in 2003,
we find completely consistent results.  In Fig.~\ref{spectrum} we show the
composite \textit{Chandra~} and \textit{INTEGRAL~} spectra.  In the energy
range of \textit{Chandra~}, both spectra A and B are plotted. Overall our
analysis is fully consistent with that of Roberts et al. (2003) both in
terms of flux level and spectral shape:  indeed our spectrum A has the same
photon index measured by the above authors\textit{\ }for the PWN and
spectrum B is a combination of their pulsar and PWN spectra. We have also attempted to estimate a cut-off energy in the 
pulsar and PWN spectra.
This was done by fitting  spectrum B  with two cut-off
power-law components plus two soft PSHOCK components. All the parameters 
concerning the PWN were frozen to the values found fitting spectrum A, with 
the exception of the cut-off energy. In this way we have been able to find a 
lower limit
to the pulsar cut-off energy at 80 keV in line with  similar results 
found by Roberts et al. (2004) using RXTE data. Unfortunately, no 
information could be gained on the PWN cut-off energy.

\section{Association with HESS J1809-193}
\label{discussion}

Recently the HESS collaboration (Aharonian et al. 2007) have reported the
detection of a new source \hbox{HESS~J1809$-$193} in this region, which they classify
as a candidate PWN.  \hbox{HESS~J1809$-$193} marginally overlaps with the location of \hbox{PSR~J1809$-$1917},
and the centroid of the HESS emission is separated from \hbox{PSR~J1811$-$1925} by
about 23 arc minutes.  Kargaltsev \& Pavlov (2007) have recently presented
results of a \chandra detection of X-ray emission from the nearby pulsar
J1809$-$1917 and have resolved its PWN. This pulsar/PWN system is situated $8\arcmin$
North of the HESS source centroid, but just within the HESS envelope. The
pulsar is moving rapidly in the direction of the HESS source, and has an
elongated PWN lying along the North-South direction, which may be explained
in terms of ram pressure generated by its fast motion through the local
interstellar medium. The compact PWN appears to be inside a more extended
region of X-ray emission, which is stretched towards the South of the
pulsar, i.e. in the direction of \hbox{HESS~1809$-$193}. No jet, which could lie
along the direction of motion and hence in the direction of the HESS source
(e.g. Ng \& Romani 2004), is visible in the \chandra images. Kargaltsev \&
Pavlov (2007) consider the possible but, as yet, uncertain association
between \hbox{PSR~J1809$-$1917} and \hbox{HESS~J1809$-$193}. Here we outline the possible, but
somewhat more uncertain association between \hbox{PSR~J1811$-$1925} and \hbox{HESS~J1809$-$193}. 

From an energetic point of view both pulsars are capable of supplying the
requisite instantaneous energy to power the HESS emission at about the 1\%
level of the spin down power (See Table 1). If we assume that the lifetime
energy output from the pulsar is a more significant factor for determining
the TeV fluxes, then for a leptonic model, we should integrate $\dot{E}$ over the
lifetime of the TeV-producing electrons, or the age of the pulsar, whichever
the shorter. An electron with energy of typically 20\,TeV is required to
generate a 1\,TeV photon by the Inverse Compton (IC) process off the Cosmic
Microwave Background (CMB) and seed photons supplied from bright infrared
emission from molecular clouds in the region. The associated electron
lifetimes are dependent on the intensity of the ambient magnetic field as well as
the rate of IC losses. During its $\sim1620$ year lifetime, assuming a constant value for $\dot{E}$, \hbox{PSR~J1811$-$1925}
is likely to release more energy than
\hbox{PSR~J1809$-$1917}, provided the local magnetic field is greater than $\sim
5-10 $\,$\mu $G.  In this case the synchroton lifetime of the TeV emitting particles is several times the age of G11.2-0.3, and therefore \hbox{PSR~J1809$-$1917} could have injected as many or more particles currently emitting at TeV energies into the surroundings than \hbox{PSR~J1811$-$1925} could have in its short lifetime.

The absence of a positive soft gamma-ray flux measurement from \hbox{PSR~J1809$-$1917} is
curious. Electron energies of typically 2 $\times $10$^{14}$\,eV in an ambient magnetic field of $\sim$10\,$\mu$G are required
to create the soft $\gamma $-rays by the synchrotron process, one order of
magnitude higher than those needed for the TeV photons from Compton
upscatters. INTEGRAL observations demonstrate that \hbox{PSR~J1811$-$1925}  does generate
soft $\gamma $-rays and hence is visibly a better accelerator of high energy
electrons. The INTEGRAL 20--100\,keV luminosity $2\sigma$ upper limit of $\sim 7%
\times 10^{33}$\,erg\,s$^{-1}$ for \hbox{PSR~J1809$-$1917} implies less than 0.4\% of the
spin down energy is converted into soft $\gamma $-rays. If the 20--100\,keV
luminosity is closer to the combined \hbox{PSR~J1809$-$1917} pulsar and PWN 1--10\,keV X-ray
luminosity of $\sim4.4\times 10^{32}$\,erg\,s$^{-1}$ measured by \textit{%
Chandra} (Kargaltsev \& Pavlov 2007), then this 20--100\,keV emission will
reduce to a value close to 0.05\% $\dot{E}$. Generally the luminosities of pulsars
associated with INTEGRAL and HESS sources share a surprisingly close fraction of
the instantaneous spin down energy (e.g Vela: INTEGRAL $\simeq $0.016\% and
HESS $\simeq $0.014\%; MSH 15-52: INTEGRAL $\simeq $1.8\% and HESS $\simeq $%
0.6\%; \hbox{PSR~J1617$-$5055}: INTEGRAL $\simeq $0.4\% and HESS $\simeq $1.2\%, Carrigan et al. 2007).

An examination of the morphological details of the two pulsars and the HESS
source completely reverses the situation. \hbox{PSR J1809$-$1917} clearly provides the
most plausible companion to \hbox{HESS J1809$-$193}. As the source names implies, \hbox{PSR
J1809$-$1917} is only 5\,pc (taking 3.5\,kpc as the distance to the pulsar) from the
centroid of the VHE source, whereas \hbox{PSR J1811$-$1925} is 29\,pc away (taking a
distance of 5\,kpc to the pulsar). The high velocity motion of \hbox{PSR~J1809$-$1917}
vectors towards the centre of the HESS source is another point in favour of an association of this pulsar 
to the TeV source.  If the spin axis and
hence possibly the putative jet are aligned with the direction of motion as discussed by Ng
\& Romani (2004), a jet may be a possible, although unproven, 
mechanism to feed high energy electrons ahead of the rapidly moving pulsar
and directly into the heart of \hbox{HESS~J1809$-$193}. The faint X-ray structure
seen in the Chandra images (Kargaltsev \& Pavlov 2007) to the south of the
pulsar may well provide, in future and deeper investigations, a firm
observational basis for this scenario. Conversely, such
evidence is lacking for \hbox{PSR J1811$-$1925}.  The jet-like structure seen in the X-ray images does
not point directly to the heart of the TeV source and there is no
observational evidence to support the passage of energetic particles through
the walls of the surrounding G11.2-0.3 supernova remnant.

\begin{table}
\caption{Characteristics of the pulsars located near to the HESS J1809-193
source. L$_{HESS}$ covers the 1--10 TeV energy range.}
\label{compare}%
\begin{tabular}{|c|c|c|c|}
\hline
Pulsar & $\dot{E}$ (erg/s) & Distance (kpc) & L$_{\mathrm{HESS}}(\% \dot{E}$)
\\ 
\hline
PSR J1811$-$1925 & 6.4$\times$ 10$^{36}$ & 5 & 0.6 \\ 
PSR J1809$-$1917 & 1.8$\times$ 10$^{36}$ & 3.5 & 1.2 \\ \hline
\end{tabular}%
\end{table}

\section{Summary \& Conclusions}
The INTEGRAL/IBIS telescope has detected soft $\gamma $-ray emission from
the site of \hbox{PSR~J1811$-$1925}. Comparison of the IBIS/ISGRI and \textit{Chandra}
spectra suggests that the soft $\gamma $-ray emission is not coming from the
SNR but originates within the complex structure associated with the pulsar
and its PWN system. The composite X-ray and soft $\gamma $-ray spectrum
indicates the pulsar itself provides a greater fraction of the emission seen
by IBIS; its spectrum is a hard power law showing no sign of a cut off up to
at least 80\,keV; the PWN contribution to the flux above 10\,keV is smaller
and well described by a steeper power law than the pulsar. 

Hard ($>2$\,keV) pulsed X-ray emission is believed to come from
the magnetosphere of the rotating neutron star and is dominated by
non-thermal radiation. Two kinds of models can explain this non-thermal
pulsed X/$\gamma $-ray emission: the polar cap model (Zhang \& Harding 2000)
and the outer gap model (Cheng, Ho \& Ruderman 1986a,b; Zhang \& Cheng
1997; Wang et al. 1998; Cheng et al. 1998).  In the outer gap model, the high energy emission would originate from the tail of 
the synchrotron component, while in the
polar cap model it would stem from curvature radiation. In the latter case a 
spectral cut-off is expected at GeV energies while in the former a break is predicted 
at lower MeV frequencies.  Either way, the spectral turnover is expected above the IBIS/ISGRI energy band, and broader spectral sampling is required to place constraints on any spectral break.  

It is interesting to note, that a fraction of the 20--100\,keV emission, 
roughly 50\%, is coming from the PWN
associated to \hbox{PSR~J1811$-$1925}; i.e. this is one of the few cases (notably 
Crab and Vela) where emission above 10\,keV is detected from a PWN.  The nature of this hard X-ray emission is not yet clear, as both the 
synchrotron and inverse Compton processes are likely emission mechanisms. 
Perhaps, the Spectral Energy Distribution (SED) of INTEGRAL detected PWN 
could be used to tackle this issue, provided that in all cases
we can separate the pulsar from the PWN as we have done here for \hbox{PSR~ 
J1811$-$1925}. Comparison with better studied objects like Crab and Vela can 
help in discriminating between competing models and provide more insight
into the physics of PWN.  To reach this objective the close relationship 
between VHE gamma-ray  emission and energetic young pulsars, and in a number 
of cases INTEGRAL sources, can be of valuable help. It is a pity that in the
particular case of \hbox{PSR~J1811$-$1925} the association with \hbox{HESS~J1809$-$193} is 
highly improbable.
\hbox{PSR~J1811$-$1925} is the most energetic young pulsar within the complex region
surrounding the TeV source, but not the only one capable of providing the 
requisite energy.  \hbox{PSR~J1809$-$1917}, on balance, provides a more plausible 
counterpart, although it is still quite possible that neither fuels the VHE
emitter, which could instead be powered by one of the supernova remnants 
which exist in the region.

\section*{Acknowledgements}

Based on observations with INTEGRAL, an ESA project with instruments and
science data centre funded by ESA member states (especially the PI
countries: Denmark, France, Germany, Italy, Switzerland, Spain), Czech
Republic and Poland, and with the participation of Russia and the USA.  University of Southampton authors acknowledge funding from the PPARC grant
PP/C000714/1. INAF/IASF--Roma and INAF/IASF--Bologna authors acknowledge
funding from ASI I/R/008/07/0.  The authors thank the reviewer, Mallory Roberts, for constructive criticism which has significantly improved the paper.

\end{document}